\documentclass[11pt, letterpaper]{elsarticle}
\usepackage{amsmath}
\usepackage{amssymb}
\usepackage{amsfonts}
\usepackage{graphicx}
\usepackage{multirow}
\usepackage{latexsym}
\usepackage{epic}
\usepackage{url}
\usepackage{soul}
\usepackage{afterpage}
\usepackage[unicode, pdftex]{hyperref}
\usepackage{multicol}
\usepackage{bbm}
\usepackage{setspace}
\usepackage{enumitem}
\usepackage{cleveref}
\usepackage[toc]{appendix}
\usepackage[norelsize,ruled, lined,linesnumbered]{algorithm2e}
\usepackage{tikz}
\usepackage{tikz,fullpage}
\usepackage{pgf}
\usetikzlibrary{arrows,automata}
\usepackage{tkz-berge}
\usepackage{amsthm}
\usepackage[symbol]{footmisc}
\usepackage[mathscr]{euscript}
\usepackage[left=1in,top=1in,right=1in,bottom=1in,nohead]{geometry}
\usepackage{empheq}
\usepackage{mathtools}
\DeclareMathOperator{\argmax}{argmax}
\DeclareMathOperator{\argmin}{argmin}

\newcommand{\PMI}{\textbf{PMI}}
\makeatletter
\newtheorem*{rep@theorem}{\rep@title}
\newcommand{\newreptheorem}[2]{%
	\newenvironment{rep#1}[1]{%
		\def\rep@title{#2 \ref{##1}}%
		\begin{rep@theorem}}%
		{\end{rep@theorem}}}
\makeatother
\SetAlgoNlRelativeSize{-1}
\SetKwFor{For}{for}{}{end}
\SetKwFor{While}{while}{}{end}
\newreptheorem{example}{Example}
\newtheorem{theorem}{Theorem}

\newtheorem{proposition}{Proposition}

\newtheorem{corollary}{Corollary}

\newtheorem{example}{Example}

\usepackage{xcolor}
\usepackage[colorinlistoftodos,prependcaption,textsize=tiny]{todonotes}
\usepackage{xargs}

\usepackage{caption}
\usepackage{subcaption}

\usepackage{lineno}
\usepackage{float}
\usepackage[section]{placeins}
\usepackage{arydshln}
\usepackage{latexsym}
\usepackage{epic}
\usepackage{amsmath,amsthm,amssymb}
\usepackage{graphicx}
\usepackage{url}
\usepackage{pgfplots}\usetikzlibrary{plotmarks}
\usepackage{soul}
\usepackage{afterpage}
\usepackage{cleveref}
\usepackage{bbm}
\usetikzlibrary{decorations.markings}
\usepackage{setspace}
\usepackage{appendix}
\usepackage{comment}
\DeclareGraphicsExtensions{.pdf,.png,.jpg,.bmp}

\allowdisplaybreaks

\begin{document}
	\begin{frontmatter}
		\title{On a class of interdiction problems with partition matroids: complexity and polynomial-time algorithms\footnote{This is the author accepted manuscript (AAM). The final published version is available at: https://doi.org/10.1287/ijoc.2024.0599}}
		
		\author[label1]{Sergey S.~Ketkov\footnote[2]{Corresponding author. Email: sergei.ketkov@business.uzh.ch; phone: +41 078 301 85 21.}}
		\author[label1]{Oleg A. Prokopyev}
		\address[label1]{Department of Business Administration, University of Zurich, Zurich, 8032, Switzerland}
		\begin{abstract}
			\onehalfspacing
			\looseness-1 In this study, we consider a class of linear matroid interdiction problems, where the feasible sets for the upper-level decision-maker (referred to as a \textit{leader}) and the lower-level decision-maker~(referred to as a \textit{follower}) are induced by two distinct partition matroids with a common weighted ground~set. Unlike classical network interdiction models where the leader is subject to a single budget constraint, in our setting, both the leader and the follower are subject to several independent \textit{capacity constraints} and engage in a zero-sum game. While the problem of finding a maximum weight independent set in a partition matroid is known to be polynomially solvable, we prove that the considered bilevel problem is \textit{NP}-hard even when the weights of ground elements are all binary. On a positive note, it is revealed that, if the number of capacity constraints is fixed for either the leader or the follower, then the considered class of bilevel problems admits several polynomial-time solution schemes. Specifically, these schemes are based on a single-level dual reformulation, a dynamic programming-based approach, and a greedy algorithm for the leader.	
		\end{abstract}
		
		
	\end{frontmatter}
	\onehalfspacing
	\section{Introduction} \label{sec: intro}
	\subsection{Motivation and background} \label{subsec: motivation}
	\looseness-1 Interdiction forms a broad class of deterministic and stochastic optimization problems, arising, for example, in the military, law-enforcement and infectious disease control contexts; see, e.g., the surveys in \citep{Kleinert2021, Smith2013, Smith2020, Wood2011} and the references therein. A classical attacker-defender network interdiction model can be viewed as a zero-sum game between two decision-makers: the upper-level decision-maker, commonly referred to as a \textit{leader} or an attacker, and the lower-level decision-maker, referred to as a \textit{follower} or a defender. The leader initiates the game by allocating limited interdiction resources to attack components of the follower’s network. The follower, in turn, observes the leader’s attack and attempts to optimize its cost function in the interdicted network. Notably, the leader is informed about the follower’s objective criterion and, therefore, it chooses attacks that maximize the cost incurred by the follower; see, e.g., \cite{Wood2011}. 
	
	\looseness-1 The majority of interdiction models in the literature focus on network interdiction, e.g., interdicting the shortest path \citep{Fulkerson1977, Israeli2002}, the minimum spanning tree \citep{Frederickson1999}, or the maximum matching~\citep{Zenklusen2015}. However, more recent studies explore interdiction problems not directly connected to graphs. These problems include knapsack interdiction \citep{Caprara2016, Fischetti2019}, matroid interdiction~\citep{Hausbrandt2024}, and interdiction for some specific classes of linear programs \citep{Chestnut2017, Dinitz2013, Yang2023}. 
	
	In this study, we focus on a new 
	class of interdiction problems, 
	where the feasible sets of both the leader and the follower are formed by \textit{partition~matroids}.
	In line with the basic matroid theory, see e.g., \citep{Oxley2022}, a general matroid~$M$ is defined as a pair $(N, \mathcal{I})$, where $N$ is a finite \textit{ground set} and~$\mathcal{I}$ is a collection of \textit{independent sets} of $N$. Specifically, collection $\mathcal{I}$ satisfies the following properties: 
	\begin{enumerate}
		\item[(\textit{i})] The empty set is independent, i.e., $\emptyset \in \mathcal{I}$.
		\item[(\textit{ii})] Every subset of an independent set is independent, i.e., for each $A' \subseteq A \subseteq N$, if $A \in \mathcal{I}$, then $A' \in \mathcal{I}$.
		\item[(\textit{iii})] If $A$ and $B$ are independent sets and $|A| < |B|$, then there exists an element $e \in B \setminus A$ such that $A \cup \{e\} \in \mathcal{I}$.
	\end{enumerate}
	
	In a partition matroid, which is a key component of our interdiction model, the ground set $N := \{1, \ldots, n\}$ is divided into $k \leq n$ distinct groups. Each group has a maximum capacity given by the largest number of elements that can be selected from this group. An independent set of a partition matroid is any subset of $N$, where the number of elements selected from each group does not exceed its maximum capacity. In particular, if $k = 1$, then the partition matroid reduces to a \textit{uniform matroid}. Other standard types of matroids include \textit{vector matroids} (sets of linearly independent vectors), \textit{graphic matroids} (sets of edges in a graph that do not contain cycles) and \textit{transversal matroids}~(sets of vertices on one side of a bipartite graph contained in a matching); we refer the reader to \cite{Oxley2022} for further details on matroids and their properties.

	
	Given two partition matroids, $X$ and $Y$, on the same weighted ground set $N$, we consider the following zero-sum game between a leader and a follower. In the first step, the leader selects an independent set~$I$ of matroid~$X$. In the second step, the follower selects an independent set $I'$ from a matroid $Y \setminus I$, which is derived by excluding the elements in $I$ from both the ground set and the independent sets of $Y$. The follower's objective is to select a set $I'$ with a maximal possible weight, whereas the leader selects~$I$ so as to minimize the weight of the follower's independent set~$I'$. Hereafter, the aforementioned bilevel model is referred to as a \textit{partition matroid interdiction}~(PMI) problem. In addition to this general definition, we provide a standard bilevel integer linear programming (bilevel ILP) formulation~of the~PMI problem in Section~\ref{subsec: problem}.


	\looseness-1 From the practical perspective, our model can be viewed as a simplification of the multidimensional knapsack interdiction problem \cite{Chen2022, Fischetti2019}, where both the leader and the follower pick items from a common item set, and are subject to multiple, potentially interdependent budget constraints.
	More specifically, in our model, both decision-makers also operate on the same ground set $N$. However, if $k_l$ and $k_f$ denote the numbers of groups in the partition matroids $X$ and $Y$, then instead of a single follower, we may consider $k_f$ independent followers, who are subject to individual capacity constraints and have their own objective criteria. On the other hand, we may assume that in the PMI problem there are $k_l$ distinct leaders with individual capacity constraints, whose goal is to minimize the total weight collected by the followers.

	At the same time, the PMI problem leverages some concepts and ideas from network interdiction models with resource sharing~\cite{Lunday2012b, Lunday2012} and Stackelberg pricing models with multiple followers~\cite{Bohnlein2020, Briest2012}. For example, in the network interdiction models \cite{Lunday2012b,Lunday2012} the leader has different types of resources and associated budget constraints, so that each arc in the network can be affected by multiple resources simultaneously. On the other hand, in Stackelberg pricing models \cite{Bohnlein2020, Briest2012} the authors consider multiple followers, e.g., rational customers, who attempt to minimize their individual costs based on the leader's prices. 
	The relationship between our model and both knapsack interdiction and Stackelberg pricing problems is explored further in Sections \ref{subsec: related literature}-\ref{subsec: approach and contributions}. 
	
	
	In short, the aim of this study is to explore the computational complexity of the PMI problem and, if possible, to demarcate between problem classes that can be solved in polynomial time and those that exhibit $NP$-hardness. Notably, we show that the problem's complexity significantly depends on the numbers of groups associated with the partition matroids $X$~and~$Y$.

	\subsection{Related literature} \label{subsec: related literature}
	
	\textbf{Interdiction models with a single budget constraint.} It can be argued that the most well-explored matroid interdiction (MI) problems involve a single budget constraint for the leader and a follower's feasible set formed by a matroid; see, e.g., \citep{Chestnut2017, Frederickson1999, Zenklusen2010}. In this setting, the leader's budget constraint can be defined by assigning a cost $c_i \in \mathbb{R}_+$ to each ground element $i \in N$ of the matroid and imposing a limited budget $b \in \mathbb{R}_{+}$. That is, the leader makes a binary interdiction decision $\mathbf{x} \in \{0, 1\}^{|N|}$ such that 
	\begin{equation} \label{eq: budget constraint}
		\sum_{i \in N} c_i x_i \leq b.
	\end{equation} 
	In particular, if $b$ is integer and $c_i = 1$ for all $i \in N$, then the leader's feasible set induced by (\ref{eq: budget constraint}) corresponds to a uniform matroid, and constraint (\ref{eq: budget constraint}) can also be referred to as a \textit{capacity}~\textit{constraint}.
	
	\looseness-1 MI problems with a budget constraint of the form (\ref{eq: budget constraint}) and a matroid at the lower level are considered, for example, in the context of the minimum spanning tree (MST) interdiction problem~\citep{Bazgan2012, Frederickson1999, Wei2021, Zenklusen2015} and the matching interdiction problem \citep{Zenklusen2010, Zenklusen2009}. Thus, it is proved in \citep{Frederickson1999} that the~MST interdiction problem is $NP$-hard, even when (\ref{eq: budget constraint}) is a capacity constraint and the edge weights are restricted to be binary. Furthermore, the problem is shown to admit a polynomial-time constant-factor approximation algorithm \citep{Zenklusen2015}. Similar complexity results, albeit with some distinctions, are obtained for the matching interdiction problem in
	the related studies~\citep{Zenklusen2010, Zenklusen2009}. 
	
	Next, Chestnut and Zenklusen \citep{Chestnut2017} consider a more general class of interdiction problems with binary objective function coefficients, where the follower's problem admits a linear programming description and satisfies some additional structural properties. 
	For this class of problems, the authors propose an efficient 2-pseudoapproximation algorithm, which can either provide a 2-approximation of the optimal solution or a solution that is at least as good as the optimal solution but with a maximum twice budget overrun. 
	In particular, the results of Chestnut and Zenklusen \citep{Chestnut2017} can be applied to a matroid interdiction problem with monotone nonnegative submodular interdiction~costs.
	
	Finally, we refer to Hausbrandt~et~al.~\citep{Hausbrandt2024}, who analyze a parametric matroid interdiction problem with a unit budget. Their problem setting assumes a finite matroid, where the weights of ground elements depend linearly on a real parameter from a predefined parameter interval. The goal is to find, for each parameter value, a single element that, being removed from the ground set, maximizes the weight of a minimum weight independent set. For this problem, Hausbrandt~et~al.~\citep{Hausbrandt2024} develop a solution algorithm, whose running time depends on different matroid
	operations and, in particular, is polynomial for graphic matroids.
	
	Another way to model the leader's budget constraint is to charge $\delta c_i$ resources to increase the weight of element $i \in N$ by $\delta$. In this case, the leader's budget constraint has the same form as~(\ref{eq: budget constraint}) 
	but requires the leader's decision variables $\mathbf{x}$ to be non-negative and continuous; see, e.g.,~\citep{Frederickson1997, Frederickson1999}. 
	
	\looseness-1 In relation to the described model, Frederickson and Solis-Oba \citep{Frederickson1997} focus on a particular class of continuous MI problems, where the feasible set of the follower is formed by an arbitrary matroid. 
	For~this problem, the authors propose a solution algorithm that achieves strong polynomial-time complexity for matroids with a polynomial-time independence test. Their approach involves, first, reducing the MI problem to a problem over an unweighted matroid and then, transforming it into the membership problem on matroid polyhedra. Furthermore, in \citep{Frederickson2006} Frederickson and Solis-Oba design more efficient algorithms for continuous MI problems with scheduling and partition matroids by taking advantage of their special structure. 
	
	\textbf{Other related models.}
	\looseness-1 One related study is that of B{\"o}hnlein and Schaudt \citep{Bohnlein2020}, who consider Stackelberg pricing problems that are based on matroids. Specifically, the authors explore a problem setting where one or multiple followers possess their own matroids with a common ground set and seek independent sets with a minimal total weight. In~particular, the ground set is divided into two groups: the first group consists of elements with fixed weights~(prices), while the second group comprises elements for which the prices are determined by the leader. B{\"o}hnlein and Schaudt~\citep{Bohnlein2020} 
	assume that the leader aims to maximize the total price of the items selected by the followers, considering uniform, partition, and laminar matroids as the followers' feasible sets. In particular, \textit{laminar matroids} can be viewed as a generalization of partition matroids, where the groups of the ground set are either disjoint or nested.
	
	The main results of \citep{Bohnlein2020} can be summarized as follows. Using dynamic programming, it~is~proved that for a single follower and all types of the follower's feasible sets outlined above, the problem of computing leader-optimal prices can be solved in polynomial time. Furthermore, it is established that the leader's problem remains polynomially solvable when there are multiple followers and the follower's feasible set is induced by a uniform matroid. In conclusion, by a reduction from the hitting set problem, it is shown that the problem with multiple followers and partition or laminar matroids is~$NP$-hard.
	
	\looseness-1 Other related problems include a knapsack interdiction problem \citep{Caprara2013, Caprara2014, Caprara2016, DeNegre2011} and more general interdiction problems with packing constraints \cite{Chen2022, Fischetti2019}. The knapsack interdiction problem is a min-max problem, where both the leader and the follower hold their own private knapsacks and choose items from a common item set. In contrast, the models in~\cite{Chen2022, Fischetti2019} consider, in particular, a multidimensional knapsack interdiction problem, where both decision-makers hold several knapsacks and, therefore, are subject to multiple budget constraints. 
	
	\looseness-1 From the theoretical computational complexity perspective, even the standard knapsack interdiction problem has a non-matroid structure and is proved to be $\Sigma^p_2$-hard \cite{Caprara2013}. In other words, this problem is located at the second level of the polynomial hierarchy and 
	there is no way~of formulating it as an integer linear program of polynomial size, unless the polynomial hierarchy collapses; see, e.g.,~\citep{Jeroslow1985}. Despite the outlined complexity issues, there exist several solution approaches to one- and multidimensional knapsack interdiction problems that either rely on cutting plane-based techniques~\cite{Caprara2016, Fischetti2019} or polynomial-time approximation algorithms~\cite{Caprara2013, Chen2022}. In particular, Chen~et~al.~\cite{Chen2022} demonstrate that the multidimensional knapsack interdiction problem with fixed numbers of budget constraints admits a constant-factor polynomial-time approximation algorithm. 
	
	\looseness-1 Finally, we refer to Shi~et~al.~\cite{Shi2023}, who consider a specific class of bilevel mixed-integer linear programming (bilevel MILP) problems, where the follower's decision variables are all binary. In this problem setting, it is assumed 
	that the follower does not solve its optimization problem to global optimality, but seeks a locally optimal solution with respect to an $m$-flip neighborhood; see, e.g.,~\cite{Orlin2004}. 
	Shi et al.~\cite{Shi2023} demonstrate
	that, for fixed $m$, the resulting problem admits a single-level MILP reformulation of polynomial size. Furthermore, if the follower’s feasible set is a matroid, then any 2-flip locally optimal solution for
	the follower is proved to be globally optimal. Put differently, for this class of bilevel problems the proposed MILP reformulation with $m \geq 2$ yields an exact solution. 
	
	\looseness-1 In conclusion, there are several bilevel optimization models in the related literature, in which both the leader and the follower construct an independent set of a matroid together; see, e.g., \citep{Beheshti2016, Buchheim2022, Gassner2009, Shi2019}. We do not provide a thorough discussion of these models because they are related to network optimization problems, such as the MST and the maximum matching problems. Moreover, the feasible sets in the aforementioned models, individually for the leader and for the follower, do not necessarily correspond to independence sets of a~matroid. 
	
	\subsection{Our approach and contributions} \label{subsec: approach and contributions}
	As outlined earlier, the main goal of this study is to demarcate between problem classes, where the PMI problem can be solved in polynomial time and those where it turns out to be $NP$-hard. In this regard, we establish the following theoretical results:
	
	\begin{itemize}
		\item By a reduction from the independent set problem, it is shown that the PMI problem is strongly $NP$-hard, if there are no restrictions on the numbers of groups, $k_l$ and $k_f$, respectively, in the partition matroids $X$ and $Y$. This result remains valid even when the weights of ground elements are all binary. 
		\item For the case where the number of follower's groups, $k_f$, is fixed, we employ a single-level dual reformulation of the PMI problem. The resulting problem is further reduced to a polynomial number of linear optimization problems over a partition matroid. Each of these problems, in turn, is efficiently solved using the well-known greedy algorithm \citep{Edmonds1971}. 
		\item For the case where the number of leader's groups, $k_l$, is fixed, we propose a polynomial-time dynamic programming (DP)-based approach. This method involves constructing an optimal leader's decision sequentially, for each subset of the ground set controlled by the~follower. 
	\end{itemize}
	
	\looseness-1 As a result, in Table \ref{tab: results} we provide a complete complexity classification for instances of the PMI problem, based on the number of ground elements $n = |N|$ and parameters $k_l$ and $k_f$. In particular, the computational complexity of all proposed algorithms falls within the XP (exponential-time parameterized) complexity class with respect to the fixed parameter; see, e.g., \cite{Downey2012}. In other words, while the PMI problem can be solved efficiently for sufficiently small values of $k_l$ or $k_f$, the runtime grows exponentially as these parameters increase. On a positive note, Table \ref{tab: results} implies that when ~$k_l$ or $k_f$ is fixed, the computational complexity of the proposed solution techniques is independent of the other, non-fixed~parameter. 
	
	\begin{table}[H]
		\centering
		\footnotesize
		\onehalfspacing
		\begin{tabular}{c | c c}
			
			& $k_f$ is fixed & $k_f$ is not fixed \\
			\hline
			& \multirow{2}{*}{\textbf{Complexity}: $O(n^{\min\{2k_l, k_f\} + 1})$} & \multirow{2}{*}{\textbf{Complexity}: $O(n^{2k_l + 1})$} \\
			$k_l$ is fixed & \multirow{2}{*}{\textbf{Solution approach}: Duality-based, DP-based} & \multirow{2}{*}{\textbf{Solution approach}: DP-based (Section \ref{subsec: dp})}\\
			&& \\
			& \multirow{2}{*}{\textbf{Complexity}: $O(n^{k_f + 1})$} & \\
			$k_l$ is not fixed & \multirow{2}{*}{\textbf{Solution approach}: Duality-based (Section \ref{subsec: duality})} & \textbf{Complexity}: $NP$-hard (Section \ref{subsec: complexity}) \\
			& & \\
		\end{tabular}
		\caption{\footnotesize The computational complexity of the PMI problem and the respective solution approaches under different assumptions on $k_l$ and $k_f$.} 
	\label{tab: results}
\end{table}

As another line of research, in Section \ref{subsec: greedy algorithm}, we analyze a simple \textit{greedy algorithm} for the leader. The algorithm starts with an empty independent set. At each step, it updates the leader's blocking decision by adding a ground element that preserves independence and maximizes the respective change in the leader's objective function value. 
It is well known that the outlined greedy algorithm provides a constant-factor approximation for the problem of \textit{maximizing} a monotone non-decreasing submodular function subject to a single capacity constraint or matroid constraints; see, e.g., \cite{Fisher1978, Nemhauser1978}. In contrast, we demonstrate that the PMI problem corresponds to \textit{minimizing} a monotone non-increasing submodular function subject to similar constraints.

Submodularity and monotonicity offer a new perspective on the PMI problem. Thus, while minimizing a submodular function with a single capacity constraint is $NP$-hard in general~\cite{Mccormick2005}, the~PMI problem with a single capacity constraint for the leader, i.e., $k_l = 1$, remains polynomially solvable due to the DP-based algorithm; recall Table \ref{tab: results}. Moreover, given that the leader's objective function is non-increasing, it is reasonable to analyze the greedy algorithm as a potential heuristic for the leader. Surprisingly, we demonstrate that, if the leader's feasible set is an arbitrary matroid and the follower has a single capacity constraint, i.e., $k_f = 1$, then the PMI problem can be solved \textit{exactly} by the greedy algorithm.
 On the negative side, we present a counterexample showing that when $k_f \geq 2$, the greedy algorithm fails to provide any constant-factor approximation for the optimal leader’s objective function value. 

In view of the discussion above, our contributions 
to the existing interdiction literature can be summarized as follows:
\begin{itemize}
	\item We propose a novel version of the matroid interdiction (MI) problem, where the feasible sets of the leader and the follower are induced by partition matroids.
	\item In contrast to the majority of MI problems with a single budget constraint \citep{Chestnut2017, Frederickson1999, Hausbrandt2024, Zenklusen2010}, our model allows for 
	several independent capacity constraints for both decision-makers. 
	\item In contrast to the models of B{\"o}hnlein and Schaudt \citep{Bohnlein2020} and Friderickson and Solis-Oba~\cite{Frederickson1997}, we assume that the leader's decision variables are binary and its feasible set is also induced by a partition matroid. 
	\item \looseness-1 Unlike the knapsack interdiction models in \citep{Caprara2013, Caprara2016}, the PMI problem is not $\Sigma^p_2$-hard and, as we demonstrate later in Section 
	\ref{subsec: duality}, admits a single-level integer linear programming reformulation of polynomial~size. 	 Furthermore, in contrast to the multidimensional knapsack interdiction model in \cite{Chen2022}, our model admits an exact polynomial-time solution algorithm even when the number of groups (or capacity constraints) is fixed only for one of the decision-makers. 
\end{itemize}

\looseness-1 The remainder of this study is organized as follows. Sections \ref{subsec: problem} and \ref{subsec: complexity} provide a bilevel ILP formulation of the partition matroid interdiction (PMI) problem and the proof of its $NP$-hardness. In Sections~\ref{subsec: duality} -- \ref{subsec: greedy algorithm}, we develop the duality-based algorithm, the DP-based algorithm, and the greedy algorithm, respectively, for several particular versions of the PMI problem. 
Finally, Section~\ref{sec: conclusion} presents our conclusions and suggests potential directions for future~research.

\textbf{Notation.} We use $\mathbb{R}_+$ to denote the set of nonnegative real numbers and $\mathbb{Z}_+$ to denote the set of nonnegative integers. Vectors are denoted by bold letters, with $\mathbf{1}$ representing the all-ones vector and $\mathbf{e}_i$ representing the $i$-th unit vector for $i \in \{1, \ldots, n\}$. We also use capital and calligraphic letters to denote sets/matroids and collection of sets, respectively. Given an arbitrary matroid $M$, a collection of its independent sets is denoted as $\mathcal{I}_M$. $\hfill\square$
\section{General case: problem formulation and its complexity} \label{sec: general case} \subsection{Integer programming formulation} \label{subsec: problem}
Formally, given two partition matroids $X$ and $Y$, we express the PMI problem described in Section \ref{subsec: motivation} as a min-max integer linear programming problem of the form: 
\begin{align*} 
	\mbox{[\PMI]: \quad } & \min_{\mathbf{x} \in \, \mathcal{I}_X } \; \boldsymbol{\beta}^\top \mathbf{y}^*(\mathbf{x}) \\
	\mbox{s.t. }	& \mathbf{y}^*(\mathbf{x}) \in \argmax_{\,\mathbf{y} \in \, \mathcal{I}_Y(\mathbf{x}) } \boldsymbol{\beta}^\top \mathbf{y}, \nonumber
\end{align*}
where $\boldsymbol{\beta} \in \mathbb{R}_+^n$ is a nonnegative weight vector, while the feasible sets of the leader and the follower, respectively, are defined as: 
\begin{subequations} 
	\begin{align} 
		& \mathcal{I}_X = \Big \{\mathbf{x} \in \{0, 1\}^n: \; \sum_{i \in L_{k} } x_i \leq l_{k} \quad \forall k \in K_l := \{1, \ldots, k_l\} \Big \} \, \mbox{ and } \label{eq: feasible set leader} \\ 
		& \mathcal{I}_Y(\mathbf{x}) = \Big \{\mathbf{y} \in \{0, 1\}^n: \; \mathbf{y} \leq \mathbf{1} - \mathbf{x}, \; \sum_{i \in F_{k} } y_i \leq f_{k} \quad \forall k \in K_f := \{1, \ldots, k_f\} \Big \}. \textcolor{black} {\label{eq: feasible set follower}}
	\end{align}
\end{subequations} 
\begin{table}
	\centering
	\footnotesize
	\doublespacing
	\begin{tabular}{c| c}
		\hline 
		Parameter(s) & Definition \\
		\hline
		$N := \{1, \ldots, n\}$ & a ground set \\
		$\boldsymbol{\beta} \in \mathbb{R}_+^n$ & a weight vector \\
		$K_l := \{1, \ldots, k_l\}$ and $K_f := \{1, \ldots, k_f\}$ & indices of the groups for the leader and the follower \\
		
		$L_{k} \subseteq N$ and $l_{k} \in \mathbb{Z}_+$, $k \in K_l$ & { leader's partition} of $N$ and { respective} budgets of the leader \\
		$F_{k} \subseteq N$ and $f_{k} \in \mathbb{Z}_+$, $k \in K_f$ & { follower's partition} of $N$ and { respective} budgets of the follower \\ 
		\hline
	\end{tabular}
	\caption{\footnotesize Summary of the notations used in formulation [\PMI].}
	\label{tab: parameters}
\end{table}
\begin{figure}
	\centering
	\begin{tikzpicture}[scale=0.7,transform shape]
		\tikzstyle{VertexStyle}=[shape = circle,thick]
		\Vertex[x=-4,y=0,L=$1$]{1}
		\Vertex[x=4,y=0,L=$5$]{5}
		\Vertex[x=-1.3,y=0.4,L=$N$]{6}
		\Vertex[x=2,y=0,L=$4$]{4}
		\tikzstyle{VertexStyle}=[shape = circle,draw]
		\Vertex[x=-1,y=-1.5,L =$F_1$]{3 }
		\Vertex[x=1,y=-2.5,L =$F_2$]{4 }
		\tikzstyle{VertexStyle}=[shape = circle,red,thick]
		\Vertex[x=-2,y=0,L=$\textbf{2}$]{2}
		\Vertex[x=0,y=0,L=$\textbf{3}$]{3}
		\tikzstyle{VertexStyle}=[shape = circle,red,draw,thick]
		\Vertex[x=-3,y=1.5,L =$L_1$]{1 }
		\Vertex[x=2,y=1.5,L =$L_2$]{2 }
		\tikzstyle{LabelStyle}=[fill=white,sloped]
		\tikzstyle{EdgeStyle}=[post, -]
		\Edge(4 )(5)
		\tikzstyle{EdgeStyle}=[post, red, -, thick]
		\Edge(1 )(2)
		\Edge(2 )(3) 
		\tikzstyle{EdgeStyle}=[post, dashed, -]
		\Edge(1 )(1)
		\Edge(4 )(4)
		\Edge(2 )(4)
		\Edge(2 )(5)
		\Edge(3 )(3)
		\Edge(3 )(2)
		\tikzstyle{EdgeStyle}=[dashed, bend left, -]
		\Edge(4 )(1)
		\draw (0,0) ellipse (5.2cm and 0.7cm);
	\end{tikzpicture}
	\caption{\footnotesize \onehalfspacing An instance of [\textbf{PMI}] with $n = 5$, $k_l = k_f = 2$, $l_1 = l_2 = f_1 = f_2 = 1$, $L_1 = \{1, 2\}$, $L_2 = \{3,4,5\}$, $F_1 = \{2, 3\}$ and $F_2 = \{1,4,5\}$. For a given $\boldsymbol{\beta} = (1, 2, 3, 4, 5)^\top$, the optimal leader's and follower's decisions, $\mathbf{x}^* = (0, 1, 1, 0, 0)^\top$ and $\mathbf{y}^*(\mathbf{x}^*) = (0, 0, 0, 0, 1)^\top$, are highlighted, respectively, in red (bold) and~black. }
	\label{fig: problem}
\end{figure}
In particular, subsets
\[L_k \subseteq N, \; k \in K_l, \mbox { and } F_{k} \subseteq N, \; k \in K_f,\] 
divide the ground set~$N = \{1, \ldots, n\}$ into $k_l$ and $k_f$ disjoint groups. In each group, the maximum capacity is bounded, respectively, by the leader's budget $l_{k} \in \mathbb{Z}_+$, $k \in K_l$, or the follower's budget~$f_{k} \in~\mathbb{Z}_+$, $k \in K_f$, where \[\sum_{k \in K_l} l_k \leq n \, \mbox{ and } \, \sum_{k \in K_f} f_{k} \leq n.\] 

In the following, whenever we refer to a partition matroid interdiction problem or~[\PMI], it is assumed that the leader's and the follower's feasible sets are given by equations~(\ref{eq: feasible set leader}) and (\ref{eq: feasible set follower}), respectively.
To clarify the problem setup, we illustrate an instance the PMI problem along with its optimal solution in Figure~\ref{fig: problem}. Finally, Table \ref{tab: parameters} provides a short summary of~the~notations~used. 


\subsection{Proof of $NP$-hardness} \label{subsec: complexity}
\looseness-1 In this section, we demonstrate that [\PMI] is strongly $NP$-hard for non-fixed $k_l$ and $k_f$, even~when the weight vector $\boldsymbol{\beta}$ is restricted to binary values, i.e., $\boldsymbol{\beta} \in \{0, 1\}^n$. To this end, we construct a polynomial-time reduction from the \textit{maximum independent set} (MIS) problem, which is known to be $NP$-hard in the strong sense \citep{Arora2009}. A decision version of the MIS problem is formulated as follows:

\begin{itemize}
	\item[$ $] [\textbf{MIS-D}]: Given an integer $q \in \mathbb{Z}_{+}$ and a graph $G = (V, E)$, where $V$ is a set of vertices and~$E$ is a set of
	edges in $G$, is there a subset $S \subseteq V$ such that no two vertices in $S$ are adjacent to each
	other in $E$, and $|S| \geq q \,$?
\end{itemize}

Given an instance of [\textbf{MIS-D}], we construct an associated instance of the partition matroid interdiction problem~[\PMI] in the following way. Assume that $n = |V|\cdot|E|$, $K_l = \{1, \ldots, |E|\}$, $K_f = \{1, \ldots |V|\}$,  $l_{k} = 1$ for $k \in K_l$ and $f_{k} = 1$ for $k \in K_f$. Also, let $V = \{v_1, \ldots, v_{|V|}\}$, $E = \{e_1, \ldots, e_{|E|}\}$ and
\[L_k = \{(v, e_k) \mid v \in V\}, \, k \in K_l, \, \mbox{ and } \, F_k = \{(v_k, e) \mid e \in E\}, \, k \in K_f.\]
That is, the leader's and the follower's feasible sets, are defined respectively, as
\begin{subequations}
	\begin{align} \label{eq: feasible sets complexity}
		& \widetilde{\mathcal{I}}_{X} = \Big\{\mathbf{x} \in \{0, 1\}^{|V|\times |E|}: \sum_{v \in V} x_{v,e} \leq 1 \quad \forall e \in E \Big\} \, \mbox{ and } \\
		& \widetilde{\mathcal{I}}_{Y}(\mathbf{x}) = \Big\{\mathbf{y} \in \{0, 1\}^{|V|\times |E|}: \; \mathbf{y} \leq 1 - \mathbf{x}, \; \sum_{e \in E} y_{v,e} \leq 1 \quad \forall v \in V \Big\}, 
	\end{align}
\end{subequations}
Furthermore, let $\widetilde{\beta}_{v, e} = 1$, if edge $e \in E$ is incident to vertex $v \in V$, and $\widetilde{\beta}_{v, e} = 0$, otherwise. 
As a result, a decision version of [\PMI] is formulated as follows:
\begin{itemize}
	\item[$ $] [\textbf{PMI-D}]: Given an integer $ \widetilde{q} = |V| - q \in \mathbb{Z}_+ $, is there a blocking decision $\mathbf{x} \in \widetilde{\mathcal{I}}_{X}$ that satisfies \[\max_{\mathbf{y} \in \, \widetilde{\mathcal{I}}_{Y}(\mathbf{x})} \boldsymbol{\widetilde{\beta}}^\top \mathbf{y} = \boldsymbol{\widetilde{\beta}}^\top \mathbf{y}^*(\mathbf{x}) \leq \widetilde{q} \,?\]
\end{itemize}
\begin{figure}[ht!]
	\centering
	\begin{subfigure}[b]{0.45\textwidth}
		\centering
		\begin{tikzpicture}[scale=1,transform shape]
			\tikzstyle{VertexStyle}=[shape = circle,red,thick,draw]
			\Vertex[x=-1,y=-1,L=$v_1$]{1}
			\Vertex[x=1,y=1,L=$v_3$]{3}
			\tikzstyle{VertexStyle}=[shape = circle,black,draw]
			\Vertex[x=-1,y=1,L=$v_2$]{2}
			\Vertex[x=1,y=-1,L=$v_4$]{4}
			\tikzstyle{EdgeStyle}=[post, -, thick]
			\Edge[label=$e_1$](1)(2)
			\Edge[label=$e_2$](2)(3)
			\Edge[label=$e_3$](3)(4)
			\Edge[label=$e_4$](4)(1)
		\end{tikzpicture}
		\caption{[\textbf{MIS-D}]}
		\label{fig: MISP-D}
	\end{subfigure}
	\hfill
	\begin{subfigure}[b]{0.45\textwidth}
		\centering
		\begin{tabular}{c| c | c | c | c}
			& $e_1$ & $e_2$ & $e_3$ & $e_4$	\\
			\hline
			$v_1$ &	\color{red} \textbf{1} \color{black} & 0 & 0 & \color{red} \textbf{1} \color{black} \\
			\hline 
			$v_2$ & 1 & 1 & 0 & 0\\
			\hline
			$v_3$ & 0 & \color{red} \textbf{1} \color{black} & \color{red} \textbf{1} \color{black} & 0\\
			\hline
			$v_4$ & 0 & 0 & 1 & 1\\
		\end{tabular}
		\caption{[\textbf{PMI-D}]}
		\label{fig: PMI-D}
	\end{subfigure}
	\caption{ A pair of instances of [\textbf{MIS-D}] and [\textbf{PMI-D}] with $|V| = |E| = 4$, $q = 2$, $\widetilde{q} = 2$ and $\widetilde{\boldsymbol{\beta}}$ defined as the incidence matrix of graph $G$. The answer to both problems is ``yes'', e.g., with the independent set and the blocking decision highlighted in red (bold).} 
	\label{fig: reduction}
\end{figure}

\color{black}
We illustrate the proposed reduction with a simple example depicted in Figure \ref{fig: reduction}. In particular, given an independence set of $G$, we construct a blocking decision $\mathbf{x} \in \widetilde{\mathcal{I}}_{X}$ in [\textbf{PMI-D}] so that it blocks the respective rows of the incidence matrix of $G$, and vice versa. The following result~holds.

\begin{theorem} \label{theorem 1} 
	Any ``yes''-instance of \upshape [\textbf{MIS-D}] \itshape corresponds to a ``yes''-instance of \upshape [\textbf{PMI-D}], \itshape and vice versa. 
	\begin{proof}
		$\Leftarrow$ If we have a ``yes'' instance of [\textbf{PMI-D}], then there exists a blocking decision $\mathbf{x} \in \widetilde{\mathcal{I}}_{X}$ such that
		\[ \widetilde{q} = |V| - q \geq \max_{\mathbf{y} \in \, \widetilde{\mathcal{I}}_{Y}(\mathbf{x})} \widetilde{\boldsymbol{\beta}}^\top \mathbf{y} = \sum_{v \in V} \max_{\mathbf{y}_v \in \{0, 1\}^{|E|}} \Big\{ \sum_{e \in E} \widetilde{\beta}_{v, e} \, y_{v, e}: \; \mathbf{y}_v \leq 1 - \mathbf{x}_v, \; \sum_{e \in E} y_{v, e} \leq 1 \Big\},\]
		\color{black}
		where $\mathbf{x}_v = (x_{v, e})_{e \in E} $ and $\mathbf{y}_v = (y_{v, e})_{e \in E} $ for $v \in V$.
		Moreover, since $\widetilde{\boldsymbol{\beta}} \in~\{0, 1\}^{|V| \times |E|}$, we conclude that for each $v \in V$ the following equality holds: 
		\[ r_v := \max_{\mathbf{y}_v \in \{0, 1\}^{|E|}} \Big\{ \sum_{e \in E} \widetilde{\beta}_{v, e}\, y_{v, e}: \; \mathbf{y}_v \leq 1 - \mathbf{x}_v, \; \sum_{e \in E} y_{v, e} \leq 1 \Big\} = \begin{cases} 0, \mbox{ if } \widetilde{\beta}_{v, e} (1 - x_{v, e}) = 0 \quad \forall e \in E, \\ 1, \mbox{ otherwise.} \end{cases} \]
		Consequently, at least $q$ values among $r_{v}$, $v \in V$, are equal to zero. 
		
		Let $S \subseteq V$, $|S| \geq q$, be a set of vertices such that $r_{v} = 0$ for each $v \in S$. We demonstrate that~$S$ is an independent set of $G$, i.e., [\textbf{MIS-D}] admits a ``yes'' instance. Assume to the contrary that there exist two vertices, $ v' \in S$ and $ v'' \in S$, that are incident to each other, i.e., there is an edge $e = (v', v'') \in E$. By~construction of $\boldsymbol{\widetilde{\beta}}$, we have $\widetilde{\beta}_{v', e} = \widetilde{\beta}_{v'', e} = 1$. However, since $\mathbf{x} \in \widetilde{\mathcal{I}}_{X}$, it can be observed that \[x_{v', e} + x_{v'', e} \leq 1,\]
		and, hence, it cannot be the case that $\widetilde{\beta}_{v', e} (1 - x_{v', e}) = 0 $ and $\widetilde{\beta}_{v'', e} (1 - x_{v'', e}) = 0 $ simultaneously. As a result, either $r_{v'} = 1$ or $r_{v''} = 1$ that contradicts the definition of $S$. 
		
		$\Rightarrow$ Suppose that we have a ``yes'' instance of [\textbf{MIS-D}], i.e., there exists an independent set $S$ in $G$ such that $|S| \geq q$. Then, any edge $e \in E$ is incident to at most one vertex from $S$, i.e.,
		\[\sum_{v \in S} \widetilde{\beta}_{v, e} \leq 1 \quad \forall e \in E.\]
		We define a blocking decision in the following way. For any $v \in V$ and $e \in E$ let
		\begin{equation} \nonumber
			x_{v, e} = \begin{cases} 
				1, \mbox{ if } \widetilde{\beta}_{v, e} = 1 \mbox{ and } v \in S, \\
				0, \mbox{ otherwise.} \\
			\end{cases} 	
		\end{equation}
		Clearly, $\mathbf{x} \in \widetilde{\mathcal{I}}_{X}$ and, furthermore,
		\[r_v = \max_{\mathbf{y}_v \in \{0, 1\}^{|E|}} \Big\{ \sum_{e \in E} \widetilde{\beta}_{v, e} \, y_{v, e}: \; \mathbf{y}_v \leq 1 - \mathbf{x}_v, \; \sum_{e \in E} y_{v, e} \leq 1 \Big\} = 0 \quad \forall v \in S.\]
		Since $r_v \in \{0, 1\}$, we conclude that 
		\[\max_{\mathbf{y} \in \, \widetilde{\mathcal{I}}_{Y}(\mathbf{x})} \boldsymbol{\widetilde{\beta}}^\top \mathbf{y} = \sum_{v \in V} r_v \leq |V| - q,\]
		which implies a ``yes'' instance of [\textbf{PMI-D}]. This observation concludes the proof. 
	\end{proof}
\end{theorem}
\color{black}
It is rather straightforward to verify that Theorem \ref{theorem 1} provides a polynomial-time reduction from [\textbf{MIS-D}] to [\textbf{PMI-D}]. Therefore, [\textbf{PMI-D}] is strongly $NP$-hard, and, furthermore, there is no fully polynomial-time approximation scheme (or FPTAS) for [\PMI], unless $P = NP$~\citep{Garey2002}. 

\section{Polynomially solvable cases} \label{sec: poly solve} 
\subsection{Duality-based algorithm} \label{subsec: duality}
In this section, we consider the partition matroid interdiction problem [\PMI], where 
the number of follower's groups, $k_f$, is fixed. 
Without loss of generality, we assume that the weight vector $\boldsymbol{\beta}$ is sorted in non-decreasing order, i.e., 
\begin{equation} \label{eq: beta sort} 
	\beta_0 \leq \beta_1 \leq \ldots \leq \beta_n. 
\end{equation}
where we set $\beta_0 := 0$. 
The following result provides an equivalent dual reformulation of [\PMI]. 

\begin{theorem} \label{theorem 2}
	The partition matroid interdiction problem \upshape [\PMI] \itshape admits the following dual reformulation: 
	
	\begin{equation} \label{eq: bilevel problem with partition follower dual}
		\min_{\mathbf{x} \in \, \mathcal{I}_X} \, 	\sum_{k \in K_f} \; \min_{j_k \in \{0, \ldots, n\}} \Big\{\beta_{j_k} f_k \; + \; \sum_{ i \in \widetilde{F}_k(j_k)} (\beta_i - \beta_{j_k})(1 - x_i) \Big\},
	\end{equation}
	where $\widetilde{F}_k(j_k) = F_k \cap \{j_k + 1, \ldots, n\}$, $k \in K_f$.
	
	\begin{proof}
		
		In the first step, we observe that the follower's problem in [\PMI] for any fixed $\mathbf{x} \in \{0, 1\}^n$ can be viewed as a linear programming problem of the~form: 
		\begin{subequations} \label{eq: follower's problem}
			\begin{align}
				& \max_\mathbf{y} \; \boldsymbol{\beta}^\top \mathbf{y} \\
				\mbox{s.t. }
				& \mathbf{0} \leq \mathbf{y} \leq \mathbf{1} - \mathbf{x} \label{cons: follower's problem 2} \\
				& \sum_{i \in F_k} y_i \leq f_k \quad \forall k \in K_f. \label{cons: follower's problem 3}
			\end{align} 
		\end{subequations}
		In particular, the integrality constraints $\mathbf{y} \in \{0, 1\}^n$ can be relaxed, since any matroid polyhedron, i.e., a convex hull of incidence vectors of its independent sets, is integral; 
		see, e.g., \citep{Edmonds2003, Schrijver2003}. 	
		Furthermore, if superscript ``$(k)$'' for $k \in K_f$ refers to the subset of indices~$F_k$, then (\ref{eq: follower's problem}) can be expressed in a block-diagonal form as:
		\begin{equation} \label{eq: follower's problem 2}
			\sum_{k \in K_f} \max_{\mathbf{y}^{\text{\tiny (\textit{k})}}} \Big\{ \boldsymbol{\beta}^{\text{\tiny (\textit{k})}\top} \mathbf{y}^{\text{\tiny (\textit{k})}}: \; \mathbf{0} \leq \mathbf{y}^{\text{\tiny (\textit{k})}} \leq \mathbf{1} - \mathbf{x}^{\text{\tiny (\textit{k})}}, \; \mathbf{1}^\top \mathbf{y}^{\text{\tiny (\textit{k})}} \leq f_k \Big\}. 
		\end{equation} 
		
		Next, we fix $\mathbf{x} \in \mathcal{I}_{X}$ and $k \in K_f$, and consider each subproblem in (\ref{eq: follower's problem 2}) given by: 
		\begin{subequations} \label{eq: follower's subproblem primal}
			\begin{align}
				& \max_{\mathbf{y}^{\text{\tiny (\textit{k})}}} \sum_{i \in F_k} \beta_i y_i \\
				\mbox{s.t. } & y_i \geq 0 \quad \forall i \in F_k \\ & y_i \leq 1 - x_i \quad \forall i \in F_k \label{cons: follower's subproblem primal 1} \\
				& \sum_{i \in F_k} y_i \leq f_k. \label{cons: follower's subproblem primal 2}
			\end{align} 
		\end{subequations}
		Let $\gamma_i \in \mathbb{R}_+$, $i \in F_k$, and $\alpha_k \in \mathbb{R}_+$ be dual variables corresponding to constraints (\ref{cons: follower's subproblem primal 1}) and (\ref{cons: follower's subproblem primal 2}), respectively. 
		Then, an equivalent dual reformulation of (\ref{eq: follower's subproblem primal}) reads as: 
		\begin{subequations} \label{eq: follower's subproblem dual}
			\begin{align}
				& \min_{\alpha_k, \boldsymbol{\gamma}^{\text{\tiny (\textit{k})}}} \Big\{\alpha_k f_k \; + \; \sum_{i \in F_k} \gamma_i (1 - x_i) \Big\} \label{eq: follower's problem dual obj}\\ 
				\mbox{s.t. } & \gamma_i \geq \beta_i - \alpha_k \quad \forall i \in F_k \\ & \gamma_i \geq 0 \quad \forall i \in F_k \\ 
				& \alpha_k \geq 0. 
			\end{align} 
		\end{subequations}
		
		In particular, we note that $\alpha_k$ can be bounded from above by $\beta_n$. Indeed, if an optimal $\alpha_k$ satisfies~$\alpha^*_k \geq \beta_n$, then, according to our assumption in (\ref{eq: beta sort}),~$\gamma^*_i = 0$ for each $i \in F_k$. In this case, the optimal objective function value of (\ref{eq: follower's subproblem dual}) satisfies \[\alpha^*_k f_k \geq \beta_n f_k,\] 
		and, hence, $\alpha_k^* = \beta_n$. 
		
		Then, we observe that objective function (\ref{eq: follower's problem dual obj}) is bilinear in the dual variables $\gamma_i$ and the upper-level decision variables $x_i$, $i \in F_k$. 
		In order to linearize the resulting single-level reformulation, we 
		divide the interval $[0, \beta_n]$ induced by constraints $\alpha_k \geq 0$ and $\alpha_k \leq \beta_n$ into $n$ disjoint intervals \[[\beta_{j}, \beta_{j + 1}], \; j \in \{0, \ldots, n - 1 \}.\] 
		\indent If $\alpha_k \in [\beta_{j_k}, \beta_{j_k + 1}]$ for some $k \in K_f$ and $j_k \in \{0, \ldots, n - 1 \}$, then, by leveraging (\ref{eq: beta sort}), we observe that the optimal value of~$\gamma_i$ for each $i \in F_k$ is given by: 
		\begin{equation} \label{eq: follower's problem optimal gamma} \nonumber
			\gamma_i^{*} = \begin{cases}
				\beta_i - \alpha_k, \mbox{ if } i \geq j_k + 1,\\
				0, \mbox{ if } i \leq j_k.	
			\end{cases}
		\end{equation}
		Therefore, for each $k \in K_f$ the dual reformulation of~(\ref{eq: follower's subproblem primal}) is expressed as:
		\begin{align} \label{eq: follower's problem dual 3}
			\min_{j_k \in \{0, \ldots, n - 1 \}} \; & \min_{\alpha_{k} \in [\beta_{j_k}, \beta_{j_k + 1}]} \Big\{ \alpha_{k} f_k \; + \; \sum_{i \in \widetilde{F}_k(j_k)} (\beta_i - \alpha_{k})(1 - x_i) \Big\},
		\end{align} 
		where $\widetilde{F}_k(j_k) := F_k \cap \{j_k + 1, \ldots, n\}$.
		Moreover, due to linearity of the objective function in (\ref{eq: follower's problem dual 3}), for any given~$j_k \in \{0, \ldots, n - 1 \}$ the optimal value of $\alpha_k$ is such that \[\alpha_k^{*} \in \{\beta_{j_k}, \beta_{j_k + 1}\}.\] 
		
		Taking into account these observations, we derive a final reformulation of~(\ref{eq: follower's problem}) given by:
		\begin{equation}\label{eq: follower's problem dual 4}
			\begin{gathered}
				\sum_{k \in K_f} \; \min_{j_k \in \{0, \ldots, n - 1\}} \; \Big\{ \min\big\{\beta_{j_k} f_k \; + \; \sum_{ i \in \widetilde{F}_k(j_k)}^{n} (\beta_i - \beta_{j_k})(1 - x_i); \; \beta_{j_k+ 1} f_k \; + \\ \sum_{ i \in \widetilde{F}_k(j_k)} (\beta_i - \beta_{j_k + 1})(1 - x_i) \big\}\Big\} = \\
				\sum_{k \in K_f} \; \min_{j_k \in \{0, \ldots, n\}} \Big\{\beta_{j_k} f_k \; + \; \sum_{ i \in \widetilde{F}_k(j_k)} (\beta_i - \beta_{j_k})(1 - x_i) \Big\}.
			\end{gathered} 
		\end{equation}
		
		\noindent Here, we exploit the fact that the overall minimum on the left-hand side of (\ref{eq: follower's problem dual 4}) is taken with respect to $2n$ terms, $n - 1$ of which are duplicated. Eventually, the dual reformulation of [\PMI] is obtained by combining the minimum over $\mathbf{x} \in \mathcal{I}_{X}$ and the right-hand side of~(\ref{eq: follower's problem dual 4}).
	\end{proof}
\end{theorem}

The first consequence of Theorem \ref{theorem 2} is that, even for non-fixed $k_l$ and $k_f$, [\PMI] with rational parameters admits a single-level \textit{integer linear programming} (ILP) reformulation of polynomial size. That is, given the dual reformulation~(\ref{eq: bilevel problem with partition follower dual}), for each $k \in K_f$, we may introduce new variables
\begin{equation} \label{eq: new variables reformulation}
	t_k = \min_{j_k \in \{0, \ldots, n\}} \Big\{\beta_{j_k} f_k \; + \; \sum_{i \in \widetilde{F}_k(j_k)} (\beta_i - \beta_{j_k})(1 - x_i) \Big\}. 
\end{equation}
If all parameters in (\ref{eq: new variables reformulation}) are rational, then, after multiplying (\ref{eq: new variables reformulation}) by a suitable constant, we may redefine $t_k$ as integer variables. Then, an equivalent ILP reformulation of (\ref{eq: bilevel problem with partition follower dual}) is obtained by linearizing the minimum on the right-hand side of (\ref{eq: new variables reformulation}). 

Secondly, by leveraging Theorem \ref{theorem 2}, we demonstrate that [\PMI] is polynomially solvable if $k_f$ is fixed. Furthermore, this result extends to a problem setting where the leader's feasible set $\mathcal{I}_{X}$ is induced by an arbitrary matroid $X$ with a polynomial-time \textit{independence test}. 

\begin{corollary} \label{corollary 1} 
	The partition matroid interdiction problem~\upshape [\PMI]~\itshape can be solved in $O(n^{k_f + 1})$ arithmetic operations. Moreover, if $\,\mathcal{I}_{X}$ is induced by an arbitrary matroid $X$, then the resulting problem can be solved in $O\big(n^{k_f + 1} \, \varphi(n) \big)$, where $\varphi(n)$ is the worst-case complexity of the independence test~$\mathbf{x} \in \mathcal{I}_{X}$ for a given $\mathbf{x} \in \{0, 1\}^n$.
	\begin{proof} If $S := \{0, \ldots, n\}^{k_f}$, then the dual reformulation~(\ref{eq: bilevel problem with partition follower dual}) can be expressed as:
		\begin{align} \label{eq: bilevel problem with partition follower dual 2} 
				& \min_{\mathbf{x} \in \, \mathcal{I}_{X}} \; 	\sum_{k \in K_f} \; \min_{j_k \in \{0, \ldots, n\}} \Big\{\beta_{j_k} f_k \; + \; \sum_{i\in \widetilde{F}_k(j_k)} (\beta_i - \beta_{j_k})(1 - x_i) \Big\} = \nonumber \\ 
				& \;\, \min_{\mathbf{x} \in \, \mathcal{I}_{X}} \; \min_{\mathbf{j} \, \in S} \;	\sum_{k \in K_f} \Big\{\beta_{j_k} f_k \; + \; \sum_{ i \in \widetilde{F}_k(j_k)} (\beta_i - \beta_{j_k})(1 - x_i) \Big\} = \\ & \; \; \min_{\mathbf{j} \, \in S} \; \min_{\mathbf{x} \in \, \mathcal{I}_{X}} \;\sum_{k \in K_f} \Big\{\beta_{j_k} f_k \; + \; \sum_{ i \in \widetilde{F}_k(j_k)} (\beta_i - \beta_{j_k})(1 - x_i) \Big\}, \nonumber
		\end{align} 
		where we use the fact that the sum of minima equals to the minimum value obtained from all possible sums of corresponding elements and the minimization can be realized in any predefined~order. 
		\looseness-1 In particular, we observe that the objective function in~(\ref{eq: bilevel problem with partition follower dual 2}) is linear and non-increasing in $\mathbf{x}$. Therefore, for any fixed realization of $\mathbf{j} \in S$, the problem on the right-hand side of~(\ref{eq: bilevel problem with partition follower dual 2}) corresponds to finding a \textit{maximum} weight independent set of matroid $X$. 
		
		Initially, we establish that, when~$X$ is a partition matroid, the abovementioned problem for fixed $\mathbf{j} \in S$ can be solved in~$O(n \log n)$ operations using the standard greedy algorithm \citep{Edmonds1971}. Thus, it requires at most~$O(n)$ operations to compute the coefficients of the objective function in~(\ref{eq: bilevel problem with partition follower dual 2}) and $O(n \log n)$ operations to sort them in non-decreasing order. After sorting, the algorithm starts with an empty independent set~$I = \emptyset$ and processes each element $i \in N$ exactly once to ensure that $I := I \cup \{i\}$ also forms an independent set of $X$. It is straightforward to verify that this independence test can be accomplished in~$O(1)$ arithmetic operations, if we keep track of the number of elements chosen from each subset~$L_k$,~$k \in K_l$. Hence, the total complexity of the greedy algorithm is determined by the sorting step and is given by $O(n \log n)$.
		
		As a result,
		since 
		\[|S| = (n + 1)^{k_f},\] 
		the total number of operations required for solving [\PMI] is of the order $O(n^{k_f+1} \log n)$. Meanwhile, the form of the objective function coefficients in (\ref{eq: bilevel problem with partition follower dual 2}) and the fact that $k_f \geq 1$ yield that the logarithmic term in the computational complexity can be eliminated by a preliminary sorting of the weight vector $\boldsymbol{\beta}$. 
		Finally, for an arbitrary matroid~$X$, instead of $O(1)$ operations for a single independence test, we need at most $O(\varphi(n))$ operations. This observation concludes the proof.
	\end{proof}
\end{corollary}
\looseness-1 We notice that for all important special cases of matroids, such as vector matroids, graphic matroids and transversal matroids, the complexity of an independence test, $\varphi(n)$, is polynomial in~$n$; see, e.g., \citep{Jensen1982}. Thus, Corollary~\ref{corollary 1} implies that there exists a rather broad class of polynomially solvable interdiction problems with the upper-level feasible set formed by a matroid. The only critical requirement for this class of problems is that the follower selects independent sets from a partition matroid with a fixed number of capacity~constraints. 

\subsection{Dynamic programming-based algorithm} \label{subsec: dp}
In this section, we consider [\PMI] 
with a fixed number of leader's groups, $k_l$. As outlined~in Section \ref{subsec: approach and contributions}, for this problem setting we design a polynomial-time DP-based algorithm.
The algorithm leverages the fact that the partition matroid interdiction problem [\PMI] can be expressed in a block-diagonal form: 
\begin{equation} \nonumber
	\min_{\mathbf{x} \in \, \mathcal{I}_{X}} \sum_{k \in K_f} \psi_k(\mathbf{x}^{\text{\tiny (\textit{k})}}), \end{equation}
where \[\psi_k(\mathbf{x}^{\text{\tiny (\textit{k})}}) := \max_{\mathbf{y}^{\text{\tiny (\textit{k})}}} \Big\{ \boldsymbol{\beta}^{\text{\tiny (\textit{k})}\top} \mathbf{y}^{\text{\tiny (\textit{k})}}: \; \mathbf{0} \leq \mathbf{y}^{\text{\tiny (\textit{k})}} \leq \mathbf{1} - \mathbf{x}^{\text{\tiny (\textit{k})}}, \; \mathbf{1}^\top \mathbf{y}^{\text{\tiny (\textit{k})}} \leq f_{k} \Big\},\]
 is the follower's profit associated with the subset $F_k$ under a blocking decision $\mathbf{x}^{\text{\tiny (\textit{k})}}$,  and the superscript~``$(k)$'' for~$k \in K_f$ refers to the subset of indices $F_k$.
Put differently, [\PMI] can be divided into~$k_f$ distinct stages, where at each stage $k \in K_f$ we need to identify an optimal blocking decision~$\mathbf{x}^{*\text{\tiny (\textit{k})}}$ in the subset $F_k$. 



\begin{algorithm}
	\DontPrintSemicolon \onehalfspacing 
	\textbf{Input:} feasible sets of the leader and the follower, respectively, $\mathcal{I}_{X}$ and $\mathcal{I}_Y(\mathbf{x})$ for $\mathbf{x} \in \mathcal{I}_{X}$.\\
	\textbf{Output:} an optimal decision, $\mathbf{x}^{*} \in \mathcal{I}_{X}$, and an optimal objective function value, $\boldsymbol{\beta}^\top \mathbf{y}^*(\mathbf{x}^*)$.\\
	 $k \longleftarrow k_f$ \; 
	\For{ $\mathbf{r}_k \in R_k $}
	{ $V_{k}(\mathbf{r}_{k}) \longleftarrow \min_{\,\mathbf{x}^{\text{\tiny (\textit{k})}} \in A^{\text{\tiny (\textit{k})}} (\mathbf{r}_{k})} \psi_k(\mathbf{x}^{\text{\tiny (\textit{k})}})$ \;
		$\tilde{\mathbf{x}}^{*\text{\tiny (\textit{k})}} (\mathbf{r}_{k}) \in \argmin_{\,\mathbf{x}^{\text{\tiny (\textit{k})}} \in A^{\text{\tiny (\textit{k})}} (\mathbf{r}_{k})} \psi_k(\mathbf{x}^{\text{\tiny (\textit{k})}})$ \;
	}
	
	\For{$k \in \{k_f - 1, \ldots, 1\}$}{
		\For{ $\mathbf{r}_k \in R_k$} {
			$V_k (\mathbf{r}_{k}) \longleftarrow \min_{\,\mathbf{x}^{\text{\tiny (\textit{k})}}} \Big\{\psi_k(\mathbf{x}^{\text{\tiny (\textit{k})}}) + V_{k + 1} \big(\boldsymbol{\theta}_k(\mathbf{r}_{k}, \mathbf{x}^{\text{\tiny (\textit{k})}})\big): \mbox{ } \mathbf{x}^{\text{\tiny (\textit{k})}} \in A^{\text{\tiny (\textit{k})}} (\mathbf{r}_{k}) \Big\}$\;
			$\tilde{\mathbf{x}}^{*\text{\tiny (\textit{k})}}(\mathbf{r}_k) \in \arg\min_{\,\mathbf{x}^{\text{\tiny (\textit{k})}}} \Big\{\psi_k(\mathbf{x}^{\text{\tiny (\textit{k})}}) + V_{k + 1} \big( \boldsymbol{\theta}_k(\mathbf{r}_{k}, \mathbf{x}^{\text{\tiny (\textit{k})}}) \big): \mbox{ } \mathbf{x}^{\text{\tiny (\textit{k})}} \in A^{\text{\tiny (\textit{k})}} (\mathbf{r}_{k}) \Big\}$\;
			
		}
	}
	$\mathbf{r}^*_{1} \longleftarrow ( l_{k'} )_{ k' \in K_l } $\;
	$\mathbf{x}^{*\text{\tiny (1)}}  \longleftarrow  \tilde{\mathbf{x}}^{*\text{\tiny (1)}}  (\mathbf{r}^*_{1} )$ \;
	\For{$k \in \{2, \ldots, k_f\}$}{ 
		$ \mathbf{r}^*_{k} \longleftarrow \boldsymbol{\theta}_{k - 1}(\mathbf{r}^*_{k - 1}, \mathbf{x}^{*\text{\tiny(\itshape{k} \upshape - 1)}})$\;
		$\mathbf{x}^{*\text{\tiny (\textit{k})}} \longleftarrow \tilde{\mathbf{x}}^{*\text{\tiny (\textit{k})}}(\mathbf{r}^*_k)$	
	}
	\Return{$\mathbf{x}^{*}$, $V_1(\mathbf{\mathbf{r}^*_1})$.}
	\caption{A DP-based algorithm for solving [\PMI].}
	\label{algorithm 1}
\end{algorithm}

 In the following, $k' \in K_l$ and $k \in K_f$ are used to denote the indices of the leader's and the follower's subsets. To implement dynamic programming, for each stage $k \in K_f$, we define a vector of \textit{states}~$\mathbf{r}_k \in \mathbb{Z}_+^{k_l}$ as~the leader's residual budgets at the beginning of stage $k$. 
Specifically, the sets of all possible states for stage $k = 1$ and the remaining stages $k \in K_f \setminus  \{1\}$ are defined, respectively,~as:
\begin{equation} \label{eq: feasible sets of states}
	 R_{1} = \Big\{ \mathbf{r}_{1}  \in \mathbb{Z}_+^{k_l}: \; r_{1, k'}  = l_{k'} \quad \forall k' \in K_l \Big\} \, \mbox{ and } \,
	R_k = \Big\{\mathbf{r}_k \in \mathbb{Z}_+^{k_l}: \; r_{k,k'} \leq l_{k'} \quad \forall k' \in K_l \Big\},
\end{equation}
 where $r_{k, k'}$ represents the residual budget of the leader related to the constraint $\sum_{i \in L_{k'} } x_i \leq l_{k'}$.  
Also, for each stage $k \in K_f$, we introduce a set of~\textit{actions} 
\begin{equation} \label{eq: feasible set for actions} 
	A^{\text{\tiny (\textit{k})}}(\mathbf{r}_k) = \Big\{ \mathbf{x}^{\text{\tiny (\textit{k})}} \in \{0, 1\}^{|F_{k}|}: \; \sum_{i \,\in F_k \cap L_{k'} } x_i \leq r_{k, k'}\; \quad \forall k' \in K_l \Big\},
\end{equation} 
which corresponds to feasible leader's decisions at stage $k$ for a given residual budget $\mathbf{r}_k \in R_k$. As a result, a new state, $\mathbf{r}_{k + 1}$, can be defined as a function of the previous state, $\mathbf{r}_k$, and the previously selected blocking decision,~$\mathbf{x}^{\text{\tiny (\textit{k})}}$, i.e.,
\begin{equation} \label{eq: transition function} 
	 \mathbf{r}_{k + 1} = \boldsymbol{\theta}_k(\mathbf{r}_{k}, \mathbf{x}^{\text{\tiny (\textit{k})}}) := \Big( r_{k, k'} - \sum_{i \in F_k \cap L_{k'}} x_i \Big)_{k' \in K_l}. 
\end{equation} 
\indent A pseudocode of the proposed DP-based algorithm is illustrated in Algorithm~\ref{algorithm 1}. Initially, at each stage $k \in \{k_f, \ldots, 1\}$, we compute a \textit{value function} $V_k(\mathbf{r}_k)$, which represents the minimum total loss the leader can achieve from a given state $\mathbf{r}_k \in R_k$ by following an optimal interdiction strategy; see lines 5 and 10 of Algorithm~\ref{algorithm 1}.  
In lines 6 and  11,  for each state $\mathbf{r}_k \in R_k$, we also identify an optimal leader's decision, $\tilde{\mathbf{x}}^{*\text{\tiny (\textit{k})}}(\mathbf{r}_k)$,  within the subset $F_k$.  It is noteworthy that, according to lines 6 and~ 11,  Algorithm \ref{algorithm 1} identifies \textit{some} optimal solution of~[\PMI] even when multiple optimal solutions exist. Finally, an optimal blocking decision,~$\mathbf{x}^*$, is restored in lines~14--19  of Algorithm \ref{algorithm 1}.

Overall, Algorithm \ref{algorithm 1} is based on a standard backward dynamic programming approach, operating over a finite state space and a finite action space, as defined by equations (\ref{eq: feasible sets of states}) and (\ref{eq: feasible set for actions}), respectively.
 In particular, the value functions in lines 5 and 10 of Algorithm \ref{algorithm 1} are computed using the standard recursive formula of dynamic programming.  Consequently, correctness  of Algorithm \ref{algorithm 1}  follows from Bellman's optimality principle \cite{Bellman1958}. The following result establishes the computational complexity of Algorithm \ref{algorithm 1}. 

\begin{theorem} \label{theorem 3}
	Algorithm~\ref{algorithm 1} solves the partition matroid interdiction problem \upshape [\PMI] \itshape in $O(n^{2k_l + 1})$ arithmetic operations, with a space complexity of $O(n^{k_l + 1})$. 	
	\begin{proof} In the first part of the proof, we demonstrate that an optimal solution of [\PMI], $\mathbf{x}^{*} \in \mathcal{I}_{X}$, satisfies certain structural properties. 
		Let the weight vector $\boldsymbol{\beta} \in \mathbb{R}^n_+$ be sorted in non-decreasing order, i.e., 
		\[0 \leq \beta_{1} \leq \ldots \leq \beta_{n}.\]
		
		We assert that, for any $k \in K_f$, $k' \in K_l$, and a pair of elements $i' \in F_{k} \cap L_{k'}$ and $i'' \in F_{k} \cap L_{k'}$ such that $i' < i''$, there exists an optimal solution $\mathbf{x}^{*} \in \mathcal{I}_{X}$ satisfying $x^*_{i'} \leq x^*_{i''}$. To this end, suppose that $x^*_{i'} = 1$ and $x^*_{i''} = 0$. Then, it is always possible to switch the values of these two elements. Indeed, the newly obtained solution remains feasible by the definition of~$\,\mathcal{I}_{X}$. Furthermore, it cannot increase the optimal leader's objective function value, $\boldsymbol{\beta}^\top \mathbf{y}^*(\mathbf{x}^*)$, as we block an element in~$F_k$ with a potentially larger weight.	
		
		Using the outlined structural properties of $\mathbf{x}^*$, we conclude that the number of candidate optimal values for~$\mathbf{x}^{*\text{\tiny (\textit{k})}}(\mathbf{r}_{k})$, $\mathbf{r}_{k}\in R_k$, can be reduced to 
		\[\prod_{k' = 1}^{k_l} \big( \min\{r_{k, k'}; |F_{k} \cap L_{k'}|\} + 1 \big).\]
		Furthermore, since $r_{k, k'} \in \{0, 1, \ldots, l_{k'}\}$ for any $k \in K_f$ and $k' \in K_l$, the total number of possible combinations of $\mathbf{r}_{k}$ and $\mathbf{x}^{*\text{\tiny (\textit{k})}}(\mathbf{r}_{k})$ can be bounded from above as:
		\begin{equation} \label{eq: number of operations for computing value function} 
			\begin{gathered}
				\sum_{r_{k, 1} = 0}^{l_1} \ldots \sum_{r_{k, k_l} = 0}^{l_{k_l}} \; \prod_{k' = 1}^{k_l} \big( \min\{r_{k, k'}; |F_{k} \cap L_{k'}|\} + 1 \big) = \prod_{k' = 1}^{k_l} \; \sum_{r_{k, k'} = 0}^{l_{k'}} \big( \min\{r_{k, k'}; |F_{k} \cap L_{k'}|\} + 1 \big) \\ \leq 
				\prod_{k' = 1}^{k_l} \; \sum_{r_{k, k'} = 0}^{l_{k'}} \big( r_{k, k'} + 1 \big) = \prod_{k' = 1}^{k_l} \frac{(l_{k'} + 2)(l_{k'} + 1)}{2} \leq \frac{1}{2^{k_l}} \Big(\frac{2n + 3k_l}{2k_l} \Big)^{2k_l}.
			\end{gathered}
		\end{equation}
		Here, the first inequality holds because $\min\{r_{k, k'}; |F_{k} \cap L_{k'}|\} \leq r_{k, k'}$. The last inequality follows from the inequality of arithmetic and geometric means and our assumption that~$\sum_{k' \in K_l} l_{k'} \leq n$. We conclude that there are at most $O(n^{2k_l})$ possible combinations of $\mathbf{r}_{k}$ and $\mathbf{x}^{*\text{\tiny (\textit{k})}}(\mathbf{r}_{k})$ for $k \in K_f$.
		
		First, in line 5 of Algorithm~\ref{algorithm 1} we compute the value function 
		\begin{equation} \label{eq: 1st Bellman function} 
			V_{ k }(\mathbf{r}_{k}) = \min_{\mathbf{x}^{\text{\tiny (\textit{k})}} \in A^{\text{\tiny (\textit{k})}} (\mathbf{r}_{k})} \Big\{\psi_k(\mathbf{x}^{\text{\tiny (\textit{k})}})\Big\}, 
		\end{equation}
		where $k = k_f$ and $\mathbf{r}_{k_f} \in R_{k_f}$. There are $O(n^{2k_l})$ possible combinations of $\mathbf{r}_{k_f}$~and~$\mathbf{x}^{*\text{\tiny (\textit{k}}_f \text{\tiny)}}(\mathbf{r}_{k_f})$ and, also, for any fixed optimal decision $\mathbf{x}^{*\text{\tiny (\textit{k}}_f \text{\tiny)}}(\mathbf{r}_{k_f})$, the objective function in (\ref{eq: 1st Bellman function}) can be computed in $O(|F_{k_f}| \log |F_{k_f}|)$ operations by the greedy algorithm. 
		Therefore, the total number of operations required to compute $V_{k_f}(\mathbf{r}_{k_f})$ is of the order \[O(n^{2k_l} |F_{k_f}| \log |F_{k_f}|).\] 
		Similar to the proof of Theorem \ref{theorem 2}, the logarithmic term can be eliminated by a preliminary sorting of the weight vector $\boldsymbol{\beta}$. 
		
		Next, for all other stages $k \in \{k_f - 1, \ldots, 1\}$ of Algorithm~\ref{algorithm 1}, the value function is given by:
		\begin{equation} \label{eq: l-th Bellman equation} \nonumber
			V_k (\mathbf{r}_{k}) = \min_{\mathbf{x}^{\text{\tiny (\textit{k})}} \in A^{\text{\tiny (\textit{k})}} (\mathbf{r}_{k})} \Big\{\psi_k(\mathbf{x}^{\text{\tiny (\textit{k})}}) + V_{k + 1} \big(\boldsymbol{\theta}_k(\mathbf{r}_{k}, \mathbf{x}^{\text{\tiny (\textit{k})}}) \big) \Big\};
		\end{equation} 
		recall line 11. 
		Analogously to the value function at stage $k_f$, computing $V_k (\mathbf{r}_{k})$ requires $O(n^{2k_l} |F_{k}|)$ operations, since the value function
		\[V_{k + 1}(\mathbf{r}_{k + 1}) = V_{k + 1} \big(\boldsymbol{\theta}_k(\mathbf{r}_{k}, \mathbf{x}^{\text{\tiny (\textit{k})}}) \big)\] for each $\mathbf{r}_{k + 1} \in R_{k + 1}$ is precomputed at stage $k + 1$. Finally, from lines $14$--$20$ of Algorithm~\ref{algorithm 1} we observe that 
		an optimal solution~$\mathbf{x}^*$ can be restored in~$O(n)$ operations. 
		
		As a result, the total number of operations to solve~[\PMI] is of the order~$O(n^{2k_l + 1})$. Furthermore, the space complexity is defined by the number of possible states and is given by~$O(n^{k_l + 1})$. This observation concludes the proof. 
	\end{proof}	
\end{theorem}	
Based on Theorems~\ref{theorem 2} and~\ref{theorem 3}, we can also recommend a particular solution approach for [\PMI] with both $k_l$ and $k_f$ being fixed. That is, if $k_f \leq 2k_l$, then we suggest using the duality-based approach. Otherwise, if the opposite inequality holds, then the DP-based approach is more efficient; recall the summary of our main results in Table~\ref{tab: results}. 

\subsection{ Greedy algorithm } \label{subsec: greedy algorithm}
In this section, we consider a general version of [\PMI] given by:
\begin{equation} \label{eq: general greedy}
	\min_{\mathbf{x} \in \, \mathcal{I}_X} \psi(\mathbf{x}),
\end{equation}
where \[\psi(\mathbf{x}) := \max_{\mathbf{y} \in \, I_{Y}(\mathbf{x})} \boldsymbol{\beta}^\top \mathbf{y},\]
and the feasible sets $\mathcal{I}_{X}$ and $\mathcal{I}_{Y}(\mathbf{x})$, $\mathbf{x} \in \mathcal{I}_{X}$, are given by equations (\ref{eq: feasible set leader}) and (\ref{eq: feasible set follower}), respectively. Given that $\mathbf{x}$ is binary, we demonstrate that the leader's objective function, $\psi(\mathbf{x})$, can be viewed as a monotone non-increasing and submodular set function. The following result holds. 
\begin{proposition} \label{proposition 1} 
	The leader's objective function, $\psi(\mathbf{x})$, satisfies the following properties:
	\begin{itemize}
		\item[\upshape(\textit{i})] $\psi(\mathbf{x})$ is monotone non-increasing, i.e., for any $\mathbf{x}' \in \{0, 1\}^n$ and $\mathbf{x}'' \in \{0, 1\}^n$, if $\mathbf{x}' \leq \mathbf{x}''$, then \[\psi(\mathbf{x}') \geq \psi(\mathbf{x}'').\] 
		\item[\upshape(\textit{ii})] $\psi(\mathbf{x})$ is submodular, i.e., for any $\mathbf{x}' \in \{0, 1\}^n$ and $\mathbf{x}'' \in \{0, 1\}^n$, if $\mathbf{x}' \leq \mathbf{x}''$ and $i \in N$ is such that $x'_i = x''_i = 0$, then
		\[\psi(\mathbf{x}' + \mathbf{e}_i) - \psi(\mathbf{x}') \geq \psi(\mathbf{x}'' + \mathbf{e}_i) - \psi(\mathbf{x}''). \] 
	\end{itemize} 
	\begin{proof} See \nameref{sec: appendix}. 
	\end{proof}
\end{proposition}

\begin{algorithm} \onehalfspacing 
	\DontPrintSemicolon
	\textbf{Input:} feasible sets of the leader and the follower, respectively, $\mathcal{I}_{X}$ and $\mathcal{I}_{Y}(\mathbf{x})$.\\ 
	\textbf{Output:} a feasible solution, $\mathbf{x}_g \in \mathcal{I}_{X}$, and the leader's objective function value,~$\psi(\mathbf{x}_g)$. \\
	$\mathbf{x} \longleftarrow \mathbf{0}$\;
	$N(\mathbf{x})$ $\longleftarrow$ $\{i \in N: x_i = 0 \, \mbox{ and } \, \mathbf{x} + \mathbf{e}_i \in \mathcal{I}_{X}\}$ \;
	\While{$N(\mathbf{x}) \neq \emptyset$} 
	{ 
		$i^* \in \argmin_{\,i \in N(\mathbf{x})} \Big(\psi(\mathbf{x} + \mathbf{e}_i) - \psi(\mathbf{x})\Big) $ \;
		$\mathbf{x} \longleftarrow \mathbf{x} + \mathbf{e}_{i^*}$ \;
		$N(\mathbf{x}) \longleftarrow \{i \in N: x_i = 0 \, \mbox{ and } \, \mathbf{x} + \mathbf{e}_i \in \mathcal{I}_{X}\}$ \;
	}
	$\mathbf{x}_g \longleftarrow \mathbf{x}$ \;
	\Return{$\mathbf{x}_g$, $\psi(\mathbf{x}_g)$.}
	\caption{A greedy algorithm for [\PMI].}
	\label{algorithm 2}
\end{algorithm}

As outlined in Section \ref{subsec: approach and contributions}, given that the leader's objective function, $\psi(\mathbf{x})$, is non-increasing, it is natural to consider a simple greedy algorithm for the leader in the context of [\PMI]; we refer to Algorithm~\ref{algorithm 2} for its pseudocode. The algorithm starts with an empty independent set by setting $\mathbf{x} = \mathbf{0}$ (line~3). In lines~5--9, we increment the current blocking decision $\mathbf{x}$, while preserving feasibility and maximizing the decrease in the leader's objective function value (ties broken arbitrarily). 
Finally, Algorithm \ref{algorithm 2} terminates once the current independent set can no longer be expanded. 

Notably, Algorithm \ref{algorithm 2} follows the structure of greedy algorithms commonly used for submodular maximization; see, e.g., \cite{Fisher1978}. The following result establishes the computational complexity of Algorithm \ref{algorithm 2}.

\begin{proposition} \label{proposition 2}
	Algorithm \ref{algorithm 2} for \upshape [\PMI] \itshape requires $O(n^2)$ operations. Moreover, if $\,\mathcal{I}_{X}$ is induced by an arbitrary matroid~$X$, then Algortihm \ref{algorithm 2} requires $O\big(n^{2} \, \varphi(n) \big)$ operations, where $\varphi(n)$ is the worst-case complexity of the independence test $\mathbf{x}\in \mathcal{I}_{X}$ for a given $\mathbf{x} \in \{0, 1\}^n$. 
	\begin{proof} See \nameref{sec: appendix}. 
	\end{proof}
\end{proposition}

Therefore, Algorithm \ref{algorithm 2} remains efficient even when the numbers of leader's and follower's groups, $k_l$ and $k_f$, are not fixed. For the subsequent result, we make the following mild assumption about the weight vector $\boldsymbol{\beta}$:
\begin{itemize}
	\item[\textbf{A1.}] The weights $\beta_i$, $i \in N$, are distinct and sorted in increasing order, i.e., $0 \leq \beta_1 < \beta_2 < \ldots < \beta_n$. 
\end{itemize} 
 Next, we demonstrate that, under Assumption \textbf{A1}, 
Algorithm \ref{algorithm 2} yields an exact solution for [\PMI] when the follower’s feasible set is defined by a uniform matroid, i.e., $k_f = 1$. This result is somewhat surprising, as [\PMI] with $k_f = 1$ constitutes a submodular minimization problem, which typically cannot be solved using a greedy algorithm. 

\begin{theorem} \label{theorem 4}
	If $k_f = 1$ and Assumption \textbf{A1} holds, then Algorithm \ref{algorithm 2} solves \upshape [\PMI] \itshape exactly. Furthermore, the result remains valid if the leader's feasible set $\mathcal{I}_{X}$ is formed by an arbitrary matroid~$X$.
	\begin{proof}
		%
		%
		If $k_f = 1$, then for any $\mathbf{x} \in \mathcal{I}_{X}$ the follower's feasible set can be expressed as: 
		\begin{equation} \label{eq: follower's feasible set greedy} \nonumber
			\mathcal{I}_{Y}(\mathbf{x}) = \Big \{\mathbf{y} \in \{0, 1\}^n: \; \mathbf{y} \leq \mathbf{1} - \mathbf{x}, \; \sum_{i = 1}^n y_i \leq f \Big \}.
		\end{equation} 
		Therefore, the optimal follower's decision for a fixed $\mathbf{x} \in \mathcal{I}_{X}$ is to select at most $f$ non-interdicted elements with the largest possible weights. In this case, the leader's objective function in [\PMI] reads~as:
		\begin{equation} \label{eq: leader's objective greedy}
			\psi(\mathbf{x}) := \boldsymbol{\beta}^\top \mathbf{y}^*(\mathbf{x}) = \sum_{t = j(\mathbf{x}, f)}^n \beta_t (1 - x_t),
		\end{equation}
		where $j(\mathbf{x}, f)$ is an index of the $f$-th non-interdicted element from the end in vector $\mathbf{x}$ and $j(\mathbf{x}, f) = 0$ if there is no such an element.
		
		Next, using equation (\ref{eq: leader's objective greedy}), we observe that, for any $\mathbf{x} \in \mathcal{I}_{X}$ and $i \in N(\mathbf{x})$,
		\begin{equation} \label{eq: delta psi}
			\psi(\mathbf{x} + \mathbf{e}_i) - \psi(\mathbf{x}) = \begin{cases}
				\beta_{j(\mathbf{x}, f + 1)} - \beta_i, \mbox{ if } i \geq j(\mathbf{x}, f), \\
				0, \mbox{ otherwise;}
			\end{cases}
		\end{equation}
		recall line 6 of Algorithm \ref{algorithm 2}.
		Let us define
		\[N_+(\mathbf{x}) := \Big\{i \in N(\mathbf{x}): \, i \geq j(\mathbf{x}, f)\Big\} \, \mbox{ and } \, N_-(\mathbf{x}) := \Big\{i \in N(\mathbf{x}): \, i < j(\mathbf{x}, f)\Big\}.\]
		Then, in the first stage, Algorithm \ref{algorithm 2} subsequently selects elements $i \in N_+(\mathbf{x})$ with the largest possible weight $\beta_i$; recall (\ref{eq: delta psi}) and Assumption \textbf{A1}. In the second stage, whenever $N_+(\mathbf{x}) = \emptyset$, the algorithm continues by expanding~$\mathbf{x}$ with elements
		$i \in N_-(\mathbf{x})$, breaking ties arbitrarily. 
		
		 It suffices to show that a solution $\mathbf{x}_g \in \mathcal{I}_{X}$ obtained by Algorithm \ref{algorithm 2} solves [\PMI] with $k_f = 1$. In this regard, we note that, by construction, $\mathbf{x}_g$ solves the following \textit{maximum} weight independent set problem:
		\begin{equation} \nonumber
			\max_{\mathbf{x} \in \, \mathcal{I}_X} \sum_{i = j(\mathbf{x}_g, f)}^n \beta_i x_i 
		\end{equation}
		and, consequently, by leveraging (\ref{eq: leader's objective greedy}), we have:
		\begin{equation} \label{min weight independent set greedy}
			\psi(\mathbf{x}_g) = \min_{\mathbf{x} \in \, \mathcal{I}_X} \sum_{i = j(\mathbf{x}_g, f)}^n \beta_i (1 - x_i). 
		\end{equation}
		
		Assume, for the sake of contradiction, that 
		\begin{equation} \label{eq: contradiction greedy}
			\psi(\mathbf{x}_g) > \min_{\mathbf{x} \in \, \mathcal{I}_X} \psi(\mathbf{x}). 
		\end{equation} 
		According to Theorem~\ref{theorem 2}, we have:
		\begin{equation} \nonumber
			\begin{gathered}
				 \min_{\mathbf{x} \in \, \mathcal{I}_X} \psi(\mathbf{x}) = 	\min_{\mathbf{x} \in \, \mathcal{I}_X} \, \min_{j \in \{0, \ldots, n\}} \Big\{\beta_{j} f \; + \; \sum_{ i = j + 1}^n (\beta_i - \beta_{j})(1 - x_i) \Big\} = \\ 
				\min_{j \in \{0, \ldots, n\}} \Big\{\beta_{j} f \; + \; \min_{\mathbf{x} \in \mathcal{I}_X} \sum_{ i = j + 1}^n (\beta_i - \beta_{j}) (1 - x_i) \Big\}. 
			\end{gathered}
		\end{equation}	
		Let $\mathbf{x}^* \in \mathcal{I}_{X}$ be a solution corresponding to the maximum weight independent set in $X$; this solution is unique by Assumption \textbf{A1}. Clearly, 
		\[\psi(\mathbf{x}_g) = \psi(\mathbf{x}^*),\]
		since the choice of elements with indices $i < j(\mathbf{x}_g, f)$ does not affect the leader's objective function value in (\ref{min weight independent set greedy}). Furthermore, by design of the greedy algorithm for matroids, $\mathbf{x}^*$ is also optimal for~each~subproblem
		
		\[\min_{\mathbf{x} \in \mathcal{I}_X} \sum_{ i = j + 1}^n (\beta_i - \beta_{j})(1 - x_i),\] 
		where $j \in \{0, \ldots, n\}$. Hence,
		\[\psi(\mathbf{x}_g) = \psi(\mathbf{x}^*) = \min_{\mathbf{x} \in \, \mathcal{I}_X} \psi(\mathbf{x}),\]
		that contradicts inequality (\ref{eq: contradiction greedy}). This observation concludes the proof. 
	\end{proof}
\end{theorem}

 Notably, the proof of Theorem \ref{theorem 4} shows that [\PMI] with $k_f = 1$ can be solved by leveraging the maximum weight independent set problem in the leader's matroid $X$. However, Algorithm~\ref{algorithm 2} leverages the structure of [\PMI] with $k_f = 1$ and, depending on the tie-breaking rule in line 6, can produce other optimal solutions that differ from the maximum weight independent set in~$X$. 

Given the abovementioned positive result about the greedy algorithm, we conclude our analysis with a negative result. That is, we show that Algorithm \ref{algorithm 2} does not provide any constant-factor approximation for [\PMI] even when $k_f = 2$ and the leader's feasible set is induced by a uniform matroid, i.e., $k_l = 1$. To establish this, let us consider the following example. 
\begin{example}
	\upshape
	Consider an instance of [\PMI] given by:
	\begin{subequations} \label{ex: example}
		\begin{align}
			\min_{\mathbf{x} \in \mathcal{I}_{X}} \, \psi(\mathbf{x}) := & \boldsymbol{\beta}^\top \mathbf{y}^*(\mathbf{x}) \\
			\text{s.t. } & \mathbf{y}^*(\mathbf{x}) \in \argmax_{\mathbf{y} \in \{0, 1\}^n} \Big\{ \boldsymbol{\beta}^\top \mathbf{y} : \mathbf{y} \leq \mathbf{1} - \mathbf{x}, \; \sum_{i \in F_k} y_i \leq f_k \quad \forall k \in \{1, 2\} \Big\},
		\end{align}
	\end{subequations}
	where $n = 5$, $f_1 = f_2 = 1$, $F_1 = \{1, 2, 3\}$ and $F_2 = \{4, 5\}$. Furthermore, let
	\[\mathcal{I}_{X} = \Big\{\mathbf{x} \in \{0, 1\}^n: \sum_{i = 1}^n x_i \leq l \Big\}\]
	with $l = 2$, and \[\beta = (1, 2, 3, M, M)^\top,\]
	where $M \in \mathbb{R}_+$ is sufficiently large.
	
	It can be verified that \[\mathbf{x}^* = (0, 0, 0, 1, 1)^\top\] is the optimal solution of (\ref{ex: example}) with the leader's objective function value $\psi(\mathbf{x}^*) = 3$. On the other hand, Algorithm 2 terminates at
	\[\mathbf{x}_g = (0, 1, 1, 0, 0)^\top,\]
	since $\psi(\mathbf{0}) = M + 3$ and the algorithm aims to achieve a maximal possible decrease in $\psi(\mathbf{x})$ at each individual step. Therefore, $\psi(\mathbf{x}_g) = M + 1$ and Algorithm \ref{algorithm 2} fails to provide any constant-factor approximation for (\ref{ex: example}). \hfill $\square$ 
\end{example}

\section{Conclusion} \label{sec: conclusion}
\looseness-1 In this study, we explore the theoretical computational complexity of a partition matroid interdiction (PMI) problem. The problem can be viewed as a zero-sum game between a leader and a follower, whose feasible sets are induced by partition matroids over the same weighted ground set. The leader blocks an independent set of its partition matroid so as to minimize the weight of the maximum weight independent set in the follower's matroid. In particular, the follower is not allowed to use the elements blocked by the leader. From a practical perspective, the~PMI problem can also be interpreted as a zero-sum game between multiple leaders and multiple independent followers. The leaders and the followers operate over two different partitions of the ground set and each decision-maker is subject to a single budget constraint with unit~costs. 

We prove that the considered PMI problem is strongly $NP$-hard in the general case and admits two polynomially solvable cases, whenever either the number of leader's or follower's capacity constraints is fixed. For the former case, we design a dynamic programming-based algorithm that subsequently selects a blocking decision of the leader for each group of elements controlled by the follower. For the latter case, we provide a single-level dual reformulation that is solved using a polynomial number of calls to the well-known greedy algorithm for matroids \citep{Edmonds1971}.

The aforementioned algorithms, being classical in nature, leverage the structure of the PMI problem at both the upper and the lower levels. Furthermore, the duality-based approach extends to a problem setting where the leader's feasible set is induced by an arbitrary matroid with a polynomial-time independence test.
Overall, our results imply a complete complexity classification for instances of the PMI problem, depending on the numbers of capacity constraints for the leader and for the follower.

As an additional interesting finding, we demonstrate that the PMI problem corresponds to minimizing a monotone non-increasing submodular function subject to matroid constraints. In~particular, it turns out that, under a mild assumption, the problem with a single capacity constraint for the follower can be solved exactly with a simple greedy algorithm for the leader. However, if the follower has more than one capacity constraint, then the greedy algorithm fails to provide any constant-factor approximation for the PMI problem. 

In conclusion, future research could focus on the following ideas. On the one hand, it is interesting to explore whether there exist polynomial-time constant-factor approximation algorithms for the considered PMI problem, given that the numbers of capacity constraints are not fixed. We believe that these algorithms could leverage the submodularity and monotonicity of the leader's objective function. However, while the problem of minimizing a submodular function without constraints can be solved in polynomial time \cite{Iwata2001, Schrijver2000}, there are only few studies that consider submodular minimization with multiple constraints; see, e.g., \cite{Iwata2009, Kamiyama2018}. 
Alternatively, potential approximation algorithms could draw on the ideas of Chen et al.~\cite{Chen2022}, who designed polynomial-time approximation algorithms for a more general multidimensional knapsack interdiction problem but with fixed numbers of budget~constraints. 

On the other hand, we note that our partition matroid interdiction problem admits a single-level integer programming reformulation of polynomial size. Therefore, standard Benders decompositon and cutting plane-based techniques can also be applied to provide a more effective exact solution algorithm for the considered bilevel problem; see, e.g., the survey in \cite{Kleinert2021} and the references therein. Finally, in view of the study by B{\"o}hnlein and Schaudt~\citep{Bohnlein2020}, it would be interesting to consider a matroid interdiction problem, where one of the feasible sets, either for the leader or the follower, is formed by a laminar matroid with a fixed number of capacity constraints. This problem is at least as hard as the problem with a partition matroid, but its complexity status remains open.

\textbf{Acknowledgments.} The authors are thankful to the associate editor and three anonymous referees for their constructive comments that allowed us to greatly improve the paper. 
\bibliographystyle{apalike}
\bibliography{bibliography}
\newpage
\section*{Appendix} \label{sec: appendix} 
\textbf{Proof of Proposition \ref{proposition 1}.}
(\textit{i}) Let the weight vector~$\boldsymbol{\beta} \in \mathbb{R}^n_{+}$ be sorted in non-decreasing order, i.e.,
\begin{equation} \label{eq: beta sort app}
	\beta_0 := 0 \leq \beta_1 \leq \ldots \leq \beta_n.
\end{equation}
First, we note that by definition
\[\psi(\mathbf{x}) = \sum_{k \in K_f} \psi_k(\mathbf{x}^{\text{\tiny (\textit{k})}}), \]
where 
\begin{equation} \label{eq: leader's functions greedy}
	\psi_k(\mathbf{x}^{\text{\tiny (\textit{k})}}) := \max_{\mathbf{y}^{\text{\tiny (\textit{k})}}} \Big\{ \boldsymbol{\beta}^{\text{\tiny (\textit{k})}\top} \mathbf{y}^{\text{\tiny (\textit{k})}}: \; \mathbf{0} \leq \mathbf{y}^{\text{\tiny (\textit{k})}} \leq \mathbf{1} - \mathbf{x}^{\text{\tiny (\textit{k})}}, \; \mathbf{1}^\top \mathbf{y}^{\text{\tiny (\textit{k})}} \leq f_{k} \Big\},
\end{equation}
and the superscript~``$(k)$'' for~$k \in K_f$ refers to the subset of indices $F_k$.
Clearly, the increase of~$\mathbf{x}$ can only reduce the follower's feasible sets in (\ref{eq: leader's functions greedy}) and, therefore, it cannot increase the leader's objective function value, $\psi(\mathbf{x})$. Hence, $\psi(\mathbf{x})$ is monotone non-increasing. 

(\textit{ii}) To prove submodularity, assume that $i \in F_k$ for some $k \in K_f$. Given a blocking decision~$\mathbf{x}'$, let $j_k(\mathbf{x}', f_k)$ be an index of the $f_k$-th non-interdicted element from the end in the subset $F_k$. If there is no such an element, let $j_k(\mathbf{x}', f_k) = 0$. We observe that:
\[\psi(\mathbf{x}' + \mathbf{e}_i) - \psi(\mathbf{x}') = \psi_k(\mathbf{x}'^{\text{\tiny (\textit{k})}} + \mathbf{e}_i) - \psi_k(\mathbf{x}'^{\text{\tiny (\textit{k})}}) = \begin{cases}
	\beta_{j_k(\mathbf{x}', f_k + 1)}-\beta_i, \mbox{ if } i \geq j_k(\mathbf{x}', f_k), \\
	0, \mbox{ otherwise.}
\end{cases}\]
Analogously,
\[\psi(\mathbf{x}'' + \mathbf{e}_i) - \psi(\mathbf{x}'') = \begin{cases}
	\beta_{j_k(\mathbf{x}'', f_k + 1)}-\beta_i, \mbox{ if } i \geq j_k(\mathbf{x}'', f_k), \\
	0, \mbox{ otherwise.}
\end{cases}
\]
Since $\mathbf{x}' \leq \mathbf{x}''$, we have 
\[j_k(\mathbf{x}'', f_k + 1) \leq j_k(\mathbf{x}', f_k + 1) \, \mbox{ and } \, j_k(\mathbf{x}'', f_k) \leq j_k(\mathbf{x}', f_k) .\]
Hence, by leveraging (\ref{eq: beta sort app}), we conclude that
\[\psi(\mathbf{x}' + \mathbf{e}_i) - \psi(\mathbf{x}') \geq \psi(\mathbf{x}'' + \mathbf{e}_i) - \psi(\mathbf{x}''), \]
and the result follows. 
\hfill $\square$ \\

\textbf{Proof of Proposition \ref{proposition 2}.}
First, we need $O(n \log n)$ operations to sort the weight vector $\boldsymbol{\beta}$ and~$O(n)$ operations to compute $\psi(\mathbf{x})$ at $\mathbf{x} = \mathbf{0}$; recall the proof of Corollary 1. 
Next, Algorithm~\ref{algorithm 2} proceeds in at most 
$n$ steps, where, at each step, we perform the following actions:
\begin{itemize}
	\item[(\textit{i})] update $N(\mathbf{x})$ by checking whether $\mathbf{x} + \mathbf{e}_i \in \mathcal{I}_{X}$ for every $i \in N$;
	\item[(\textit{ii})] compute $\psi(\mathbf{x} + \mathbf{e}_i) - \psi(\mathbf{x})$ for every $i \in N(\mathbf{x})$. 
\end{itemize}
Each independence test takes $O(1)$ and $O(\varphi(n))$ operations for the partition and arbitrary matroids, respectively. Furthermore, following the proof of Proposition \ref{proposition 1}, the respective change in the leader's objective function value can be computed in $O(1)$ operations, if we compute the indices $j_k(\mathbf{x}, f_k)$ and $j_k(\mathbf{x}, f_k + 1)$ for each $k \in K_f$ at the beginning of each step. The indices can be computed in $O(n)$ operations and, hence, the total number of arithmetic operations is given by $O(n^2)$ and $O(n^2 \varphi(n))$ for the partition and arbitrary matroid, respectively. This observation concludes the~proof. \hfill $\square$

\end{document}